\documentclass[preprint,12pt]{elsarticle}

\usepackage{amssymb}
\usepackage{amsmath}
\usepackage{graphicx}
\usepackage{hyperref}

\usepackage{subcaption}
\usepackage{algorithm}
\usepackage{algpseudocode}

\journal{Mathematical Biosciences}

\begin{document}

\begin{frontmatter}



\title{Stochastic Model of siRNA Endosomal Escape Mediated by Fusogenic Peptides}


\author[inst1]{Nisha Yadav}
\author[inst2]{Jessica Boulos}
\author[inst2]{Angela Alexander-Bryant}
\author[inst1]{Keisha Cook}

\affiliation[inst1]{organization={School of Mathematical and Statistical Sciences, Clemson University},
            addressline={220 Parkway Drive}, 
            city={Clemson},
            postcode={29634}, 
            state={South Carolina},
            country={USA}}
            
\affiliation[inst2]{organization={Department of Bioengineering, Clemson University},
            addressline={118 Engineering Service Drive}, 
            city={Clemson},
            postcode={29634}, 
            state={South Carolina},
            country={USA}}

 
 


\begin{abstract}
Gene silencing via small interfering RNA (siRNA) represents a transformative tool in cancer therapy, offering specificity and reduced off-target effects compared to conventional treatments. A crucial step in siRNA-based therapies is endosomal escape, the release of siRNA from endosomes into the cytoplasm. Quantifying endosomal escape is challenging due to the dynamic nature of the process and limitations in imaging and analytical techniques. Traditional methods often rely on fluorescence intensity measurements or manual image processing, which are time-intensive and fail to capture continuous dynamics. 
This paper presents a novel computational framework that integrates automated image processing to analyze time-lapse fluorescent microscopy data of endosomal escape, hierarchical Bayesian inference, and stochastic simulations. Our method employs image segmentation techniques such as binary masks, Gaussian filters, and multichannel color quantification to extract precise spatial and temporal data from microscopy images. Using a hierarchical Bayesian approach, we estimate the parameters of a compartmental model that describes endosomal escape dynamics, accounting for variability over time. These parameters inform a Gillespie stochastic simulation algorithm, ensuring realistic simulations of siRNA release events over time.
By combining these techniques, our framework provides a scalable and reproducible method for quantifying endosomal escape. The model captures uncertainty and variability in parameter estimation, and endosomal escape dynamics. Additionally, synthetic data generation allows researchers to validate experimental findings and explore alternative conditions without extensive laboratory work. This integrated approach not only improves the accuracy of endosomal escape quantification but also provides predictive insights for optimizing siRNA delivery systems and advancing gene therapy research.

\end{abstract}


\begin{highlights}
\item Exact quantification of endosomal escape.
\item Stochastic simulation model of endosomal escape.
\end{highlights}


\end{frontmatter}



\section{Introduction}
\label{sec:Introduction}

\subsection{Biological Framework}
Among the many gene therapy tools developed for RNA interference \cite{Rossi2007,Alexander-Bryant20131,pharmaceutics14071344}, small interfering RNA (siRNA) is a powerful tool for gene silencing of target messenger RNA in cancer cells \cite{KalogerosEtAl,Teng2022,ALEXANDERBRYANT2017123,QIN2013159,Karagiannis,KIM2010335,ozcan2015preclinical,GOLDBERG201395,leng2009advances,Zou2023}; oral cancer, brain cancer, ovarian cancer, breast cancer, and more. siRNA therapies can be designed to target specific cells, reducing unintended impacts on healthy cells and tissues \cite{ozcan2015preclinical,Yale2012,Goyal2022,he2023nanotechnology}. In cancer studies where drug resistance prevents the efficacy of cancer treatments, siRNA can target and silence genes responsible for this resistance, making existing treatments more effective. Effective gene silencing relies on endosomal escape, the release of siRNA from the endosome to the cytosol. Consequently, successful delivery of siRNA  allows for the targeting and silencing of specific genes that are causing or contributing to a disease \cite{ALEXANDERBRYANT2017123,SamecEtAl,SamecEtAl2,TAI2017157,OLIVEIRA2007211,CUMMINGS201992,Cantini}. siRNA cancer therapies open up new possibilities for treating various conditions that were previously difficult to manage with conventional drugs.

Nanotechnology-based approaches have shown promise in the targeted delivery of siRNA for cancer treatment. However, siRNA delivery faces challenges such as poor intracellular trafficking, degradation, off-target effects, and toxicity. Many types of nanocarriers, including liposomes, dendrimers, and micelle-based nanovectors, have been explored for siRNA delivery. With a focus on improving permeation into cells and reducing off-target effects, the developed nanocarriers overcome such limitations and enhance siRNA delivery to tumor sites \cite{pharmaceutics14071344,Goyal2022,SamecEtAl,SamecEtAl2,he2023nanotechnology, Yadav2022,martens2014intracellular, SmithEtAl2019}. Additionally, combining siRNA with anti-cancer drugs can have synergistic effects, making them a significant tool in cancer therapy \cite{Zou2023}. Overall, nanotechnology-based siRNA delivery holds prospects for improving cancer treatment, but challenges such as delivery optimization and overcoming resistance need to be addressed.

\begin{figure}[h]
    \centering
    \includegraphics[width=1\linewidth]{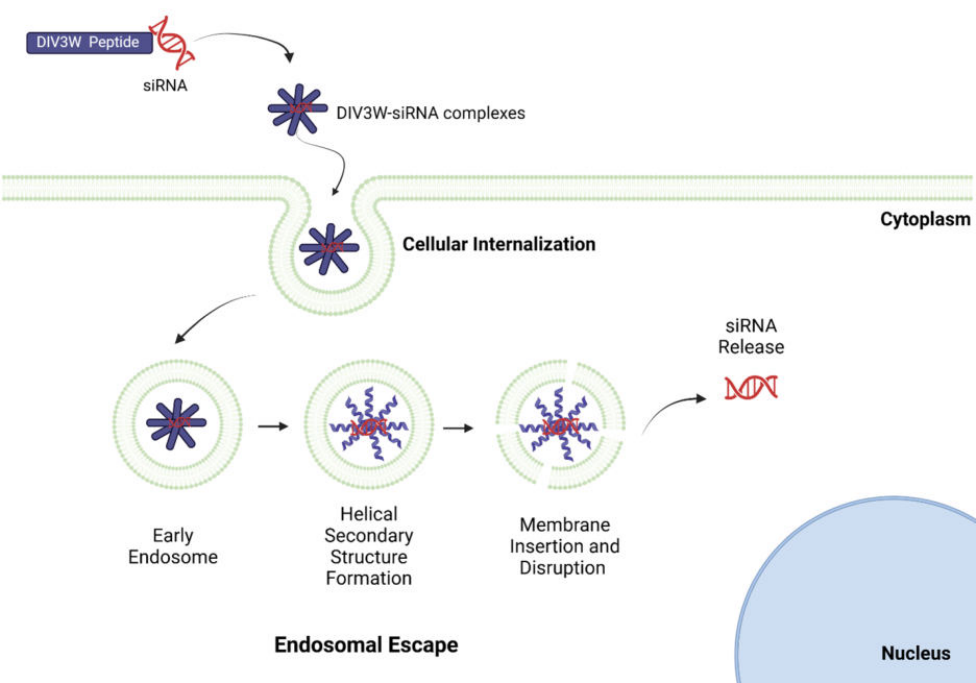}
    \caption{Diagram illustrating the biological process of endosomal escape. The DIV3W peptides facilitate internalization of DIV3W-siRNA complexes. After internalization, the complexes are encapsulated by endosomes and eventually released into the cytosol; i.e. endosomal escape. }
    \label{fig:diagram}
\end{figure}

Peptides are utilized as nanocarriers for siRNA delivery due to their cytocompatibility, affordability, and tunability for specific applications \cite{samec2022peptide}. These peptides offer functional advantages, enabling them to overcome various barriers associated with delivery, including a significant delivery barrier within gene therapy treatment: endosomal entrapment \cite{pei2018overcoming}. Typically, nanocarriers and their cargo are internalized into the cell through endocytosis, enclosed in an endosome vesicle. If the cargo remains trapped within the endosome, it will be cycled out of the cell, preventing effective delivery. Therefore, delivery systems must promote endosomal escape to release their cargo into the cytoplasm, where it can be bioactive \cite{degors2019carriers}. Fusogenic peptides, a class of stimuli-responsive sequences, are pH-dependent and can facilitate endosomal escape in response to the acidic pH within the endosome \cite{cantini2013fusogenic,oliveira2007fusogenic}. After endocytosis, these peptides undergo a conformational change, typically to a helical secondary structure, allowing them to insert into the endosomal membrane to enable and promote cargo release \cite{cantini2013fusogenic,oliveira2007fusogenic}. Figure~\ref{fig:diagram} depicts the endosomal escape process. Previous research has characterized the DIV3W peptide as an established fusogenic peptide that promotes endosomal escape of siRNA in ovarian cancer cells, facilitating oncogene silencing both in vitro and in vivo \cite{samec2022fusogenic}. Fluorescent microscopy confirmed that DIV3W internalization occurs through endocytosis, and it enables escape, demonstrated through the investigation of colocalization of tagged siRNA and endosomes. However, this data was collected at a 4-hour endpoint time frame and lacked quantification analysis.

\subsection{Existing Endosomal Escape Quantification Methods}

In many studies of endosomal escape \cite{PEI2019273, OLIVEIRA2007211, Erazo-Oliveras, Nishimura2014, HOU2015931,  Meng201774,LonnEtAl,D0NA00454E, Hausig-Punke,  Kongkatigumjorn,klipp2023get}, the verification and quantification of successful escape relies on the fluorescence intensities of labeled nanoparticles or organelles, \textit{in vivo} and \textit{in vitro}. There has been a focus on improving these techniques by improving the fluorescent methods,  assays, agents, and quantities used for fluorescent labeling \cite{dixon2016nanoluc,peier2021nanoclick, gilleron2013image}. In \cite{Rietz}, fluorescence intensities of single vesicles were measured over 24 hours to quantify drug-induced endosomal escape of siRNA, as well as the associated knockdown of genes expressed in tumor cells. In \cite{gilleron2013image}, \textit{in vitro} experiments measured the fluorescence intensity of the uptake process, endosomal escape, and delivery of siRNA over 4 hour time periods. Notably, the authors in \cite{HedlundEtAl}, use cytosol masks, cell masks, and signal-to-noise decreasing image filters to quantify the siRNA fluorescence intensities measured in individual tracked cells over 700 minutes, where one image was captured every 5 minutes. Based on the siRNA release properties, constant release, and sequential release, CellProfiler \cite{10.1093/bioinformatics/btr095} was used to quantify the fluorescence in each cell over time. The model for the increase in fluorescence intensity over time was determined by fitting a curve through the acquired quantities over time. The authors in \cite{munson2021high} use a variety of assays to measure nanoparticle uptake (occurring after 15-30 minutes), endosomal escape (occurring after about 1 hour), and delivery expression up to 4 hours using fluorescence intensities, where images were captured every 15 minutes. These fluorescence intensity quantification methods are inherently subjective and often lack consistent, reproducible results. In contrast, our approach provides precise detection of individual fluorescence-labeled objects, significantly minimizing subjectivity.

\subsection{Image Processing Tools}
Image segmentation techniques are used to partition images into categories based on the characteristics of their pixels. These techniques can be used to track the movement of clusters of cells \cite{ScherrEtAl2020}, track cell division of a group of neighboring cells \cite{PadovaniEtAl}, study the shape or type of cell \cite{HIRLING2023742}, locate cell nuclei \cite{Long2020}, and count the number of cells present in a single, two or three dimensional microscopy image \cite{DimopoulosEtAl}. Thresholding is an important feature of image segmentation, which focuses on separating the foreground objects from the background of an image and separating the foreground objects by their pixel characteristics. The OTSU method \cite{Otsu1979ATS} is a widely used technique for automatic image thresholding. It maximizes the variance between the foreground and background classes. Specifically, the OTSU method calculates the optimal threshold value by minimizing the variance within the class (the variance of each class) or equivalently maximizing the variance between classes. This is achieved through an exhaustive search over all possible threshold values to find the one that best separates the two classes. 

\section{Endosomal Escape Quantification Method Using Image Processing Tools\label{sec:quant}}

Instead of relying on the fluorescent intensity level of labeled siRNA and endosomes, we aim to quantify each labeled object present in fluorescent multichannel cellular images that are captured over multiple hours. In previous studies \cite{ SmithEtAl2019, BeachEtAl, GRAU2024102506,  TeoEtAl}, the authors show that each object can be identified by an image color channel. In some cases, only two color channels are used to differentiate between the nucleus and fluorescent organelles or particles, for which a spot count is used to quantify the amount of endosomal escape in an image. In other work \cite{Kongkatigumjorn,paramasivam2021endosomal,yazdi2024vivo}, colocalization, the step at which particles or siRNA are not yet released from the endosomes or lysosomes, is shown as the overlap of multiple color channels when there exists three or more color channels. In either case, the authors rely on images taken at time steps, 0 hours, 1 hour, 2 hours, and 4 hours which does not capture the continuous release from the endosomes between those time steps. In \cite{wittrup2015visualizing}, the authors quantify endosomal escape by fluorescence intensity over a time-lapse. We follow the concept of three color channels in our categorization of the stages of endosomal escape over a continuous time-lapse of fluorescent images. 

Our approach significantly advances the study of endosomal escape by providing a comprehensive framework for data acquisition, image analysis, and time-lapse visualization. Unlike \cite{wittrup2015visualizing}, which relies on discrete fluorescence intensity measurements prone to noise, we utilize high-resolution time-lapse imaging over 4 hours, offering continuous tracking of endosomal escape dynamics. Our automated image analysis pipeline efficiently processes raw fluorescent microscopy data, removing background noise and floating siRNA to ensure accurate, reproducible, and reliable quantification. This streamlined approach not only reduces labor-intensive manual processing time but also enhances reproducibility. We incorporate stochastic modeling to reflect the probabilistic nature of events. Additionally, synthetic data generation addresses data scarcity, ensuring robust validation of experimental results. This combination of advanced techniques ensures a more detailed, scalable, and predictive analysis compared to traditional fluorescence-based methods.

In this paper, we introduce a novel model designed for efficient image analysis and synthetic data generation to address the scarcity of experimental data in studying endosomal escape dynamics over time. The raw data is comprised of fluorescent microscopy images, which are inherently noisy because of background fluorescence and floating siRNA outside cells (as this is live-cell imaging). Our model executes the following steps:

\begin{enumerate}
    \item \textbf{Background Noise Removal}: Eliminate unwanted fluorescence signals that obscure cellular information.
    \item \textbf{Floating siRNA Removal}: Remove siRNA signals outside of cells, isolating intracellular activity.
    \item \textbf{Information Extraction}: Utilize labeled object intensity data from red, green, and blue channels to quantify key metrics related to endosomal escape.
    \item \textbf{Synthetic Data Generation}: Generate synthetic data that mimic experimental data, enabling robust studies even with limited lab-generated data.
\end{enumerate}

The first two steps, background noise removal and floating siRNA removal, are the most labor-intensive when performed manually in a lab setting. For example:

\begin{itemize}
    \item \textbf{Step 1}: Manually editing background noise in 40 images requires approximately 4 to 4.5 hours.
    \item \textbf{Step 2}: Manually removing floating siRNA requires an additional 5 to 6 hours.
\end{itemize}

\begin{figure}[h]
    \centering
    \includegraphics[width=0.75\linewidth]{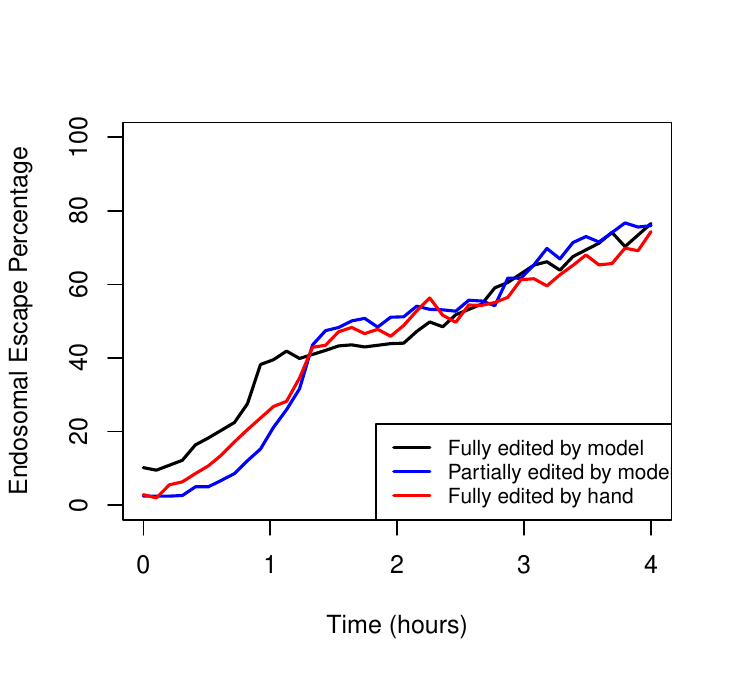}
    \caption{The percentage of endosomal escape (siRNA released from the endosomes) captured over 4 hours. Using raw data we compute the percentage using the three methods. The \textit{black} curve shows endosomal escape computed using the model developed in this paper. The \textit{blue} curve is computed by manually editing the images before using the methods described in Section \ref{sec:methods} to find the cell boundaries. The \textit{red} curve is determined by manipulating the image files by hand.}
    \label{fig:data_compare}
\end{figure}

In contrast, our model significantly reduces this time burden, as illustrated in Figure~\ref{fig:data_compare}:

\begin{itemize}
    \item The red curve represents experiments where both steps are performed manually, requiring \textbf{9–11 hours}.
    \item The blue curve shows experiments where step 1 is performed manually and step 2 is automated, reducing the time to \textbf{4–4.5 hours}.
    \item The black curve represents experiments where both steps are fully automated by the model, completing the entire process in just \textbf{50–55 seconds}.
\end{itemize}

This efficiency allows experimentalists to save significant time in image analysis while maintaining data quality. Additionally, the ability to generate synthetic data further addresses the challenges of data scarcity in experimental studies, enabling a more comprehensive analysis of endosomal escape dynamics.

Compartmental models have been used in many applications, such as epidemiology, pharmacokinetics, systems biology, and neuroscience \cite{ENDERLE2012359,brauer2008compartmental,bassingthwaighte2012compartmental}. In this work, we focus on representing the three main stages of endosomal escape; 1) endocytosis (uptake), 2) endosomal entrapment, and 3) endosomal escape \cite{SmithEtAl2019} as a compartmental model, a parameterized system of equations used to model interactions between compartments over time. We use the compartmental model to generate synthetic data of endosomal escape over time by way of the stochastic simulation algorithm \cite{Gillespie1977,Gillespie1976}, allowing us to capture the stochastic nature of fluorescent siRNA release over a 4-hour period. Additionally, we provide a method to validate the accuracy of the synthetic data by using error optimization and sensitivity analysis for measuring the plausibility of the rate parameters associated with the population fluctuations for each model compartment.  

\section{Experimental Methods}
\label{sec:experimental_methods}

\subsection{Cell Culture}
\label{Culture}

The OVCAR-3 ovarian cancer cell line was obtained from American Type Culture Collection (ATCC, Manassas, VA, USA) and cultured in RPMI Media supplemented with 20\% fetal bovine serum, 1\% penicillin/streptomycin antibiotic solution, and 0.1\% insulin (Corning, Corning, NY, USA). All cultures were incubated at $37^{\circ}$C and 5\% CO$_2$.

\subsection{Peptide Synthesis}
\label{Synthesis}

DIV3W peptide (WEADIVADIVWDIVADIVAGGG-(d)RRRRRRRRR) was synthesized, purified ($\geq 95$\% purity), and analyzed via high-performance liquid chromatography by Lifetein (Lifetein LLC, Somerset, NJ, USA).  Lyophilized peptide was resuspended in ultrapure $H_{2}O$ to a 10 mg/mL (microgram per milliliter) stock solution. Working solutions were made at 1 mg/mL with a $1:10$ dilution of stock peptide and RNase-free water. All peptide solutions were stored at $-80^{\circ}$C.

\subsection{Peptide-siRNA Complex Formation}
\label{Complex}

All peptides were complexed with 10 $\mu$M (micrometer) siGENOME non-targeting siRNA\#5 (siNT\#5)-Cy5 fluorescently tagged (GE Healthcare Dharmacon, Lafayette, CO, USA) through electrostatic complexation in RNase-free water for 20 minutes at room temperature. Peptide complexes were made at an 80:1 N:P (where N is the number of amino groups in each peptide and P is the number of phosphate groups in each siRNA) ratio to perform \textit{in vitro} experiments to investigate the dynamics of endosomal escape. For each experiment, a concentration of 1.2 $\mu L$ (microliters) of siRNA was added with a concentration of 8.27~$\mu L$ of peptides to prepare the samples. All results and analyses presented in this paper are based on this specific ratio. 

\begin{figure*}[h]
     \centering
     \begin{subfigure}[b]{0.3\textwidth}
         \centering
         \includegraphics[width=\textwidth]{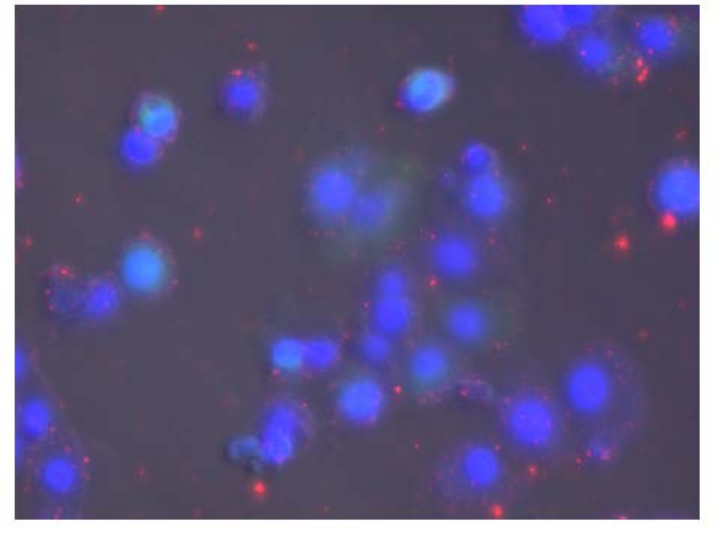}
         \caption{Original raw microscopy image.}
         \label{fig:raw_image}
     \end{subfigure}
     \begin{subfigure}[b]{0.3\textwidth}
         \centering
         \includegraphics[width=\textwidth]{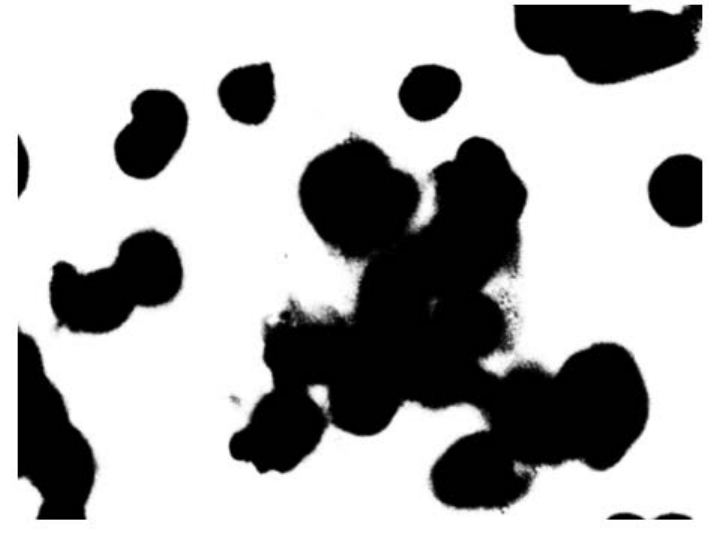}
         \caption{Mask defined by thresholding.}
         \label{fig:mask}
     \end{subfigure}
     \begin{subfigure}[b]{0.3\textwidth}
         \centering
         \includegraphics[width=\textwidth]{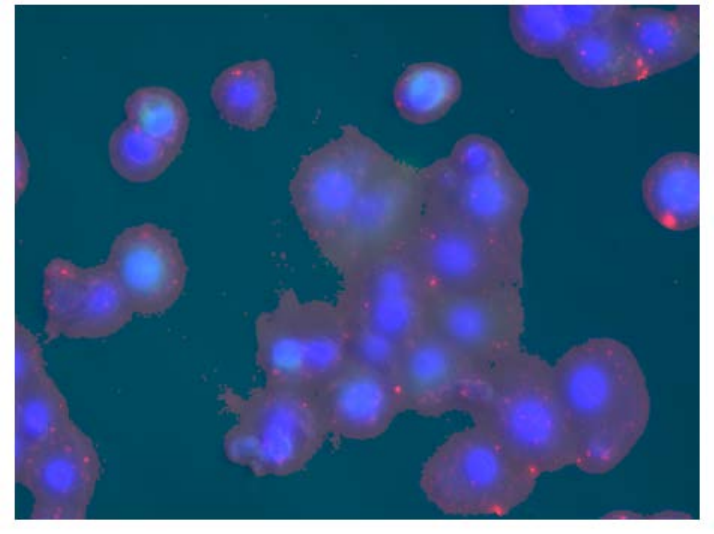}
         \caption{Microscopy image with floating siRNA removed.}
         \label{fig:remove_sirna}
     \end{subfigure}
        \caption{Fluorescent images of siRNA (red), nuclei (blue), and endosomes (green). We apply a mask to the microscopy image to remove floating siRNA (siRNA not contained within the cell membranes.). (a) The original raw fluorescent microscopy image captured at time t = 4 hours. (b) User-defined mask using thresholding technique in Section \ref{sec:imageanalysis}. (c) Fluorescent microscopy image removed with mask overlay. Floating siRNA are removed from the areas outside of the masked regions. Note that while the threshold finds very small areas that are mistaken as cells, the large areas are correct. The small areas are so small that they are negligible when it comes to counting the siRNA present inside of the boundary areas.}
        \label{fig:maskingtoremovefloatingsirna}
\end{figure*}

\section{Live Cell Imaging}
\label{livecellimaging}

OVCAR-3 cells were seeded at 10,000 cells/well and cultured in a 96-well plate overnight. The next day, fresh media was added to the cells, along with 15 $uL$ of CellLight Early Endosome-GFP-BacMam 2.0, to label the endosomes fluorescently (Thermo Fisher, Waltham, MA, USA). After overnight incubation, the media was changed and BacMam 2.0 enhancer was added for 90 minutes, media was changed, and cells were left to incubate overnight. All incubation steps were conducted at $37^{\circ}$C and in 5\% CO$_2$. The next day, DIV3W peptide was complexed with siNT\#5-Cy5 at an 80:1 ratio and Hoescht 33342 stain was added to the cells to stain the nuclei. Complexes and nuclei stain were added in media with 10\% fetal bovine serum, and live cells were imaged for 4 hours using the Stage Incubator of the BZ X800 at 40X magnification (Keyence, Itasca, IL, USA). The imaging setup included z-stack acquisition, which captures a series of optical slices through the sample at different focal planes. This approach ensures that only in-focus structures within the cell or nucleus are included in the analysis, effectively eliminating artifacts from out of focus siRNA above the cells or nuclei. Furthermore, the resolution provided by the BZ-X800 microscope is sufficiently high to allow accurate pixel level quantification of fluorescent intensity. We began capturing Images 30 minutes after complexes were added and at every 5 minutes within this time frame. 

\begin{figure*}[h]
     \centering
     \begin{subfigure}[b]{0.3\textwidth}
         \centering
            \includegraphics[width=\textwidth]{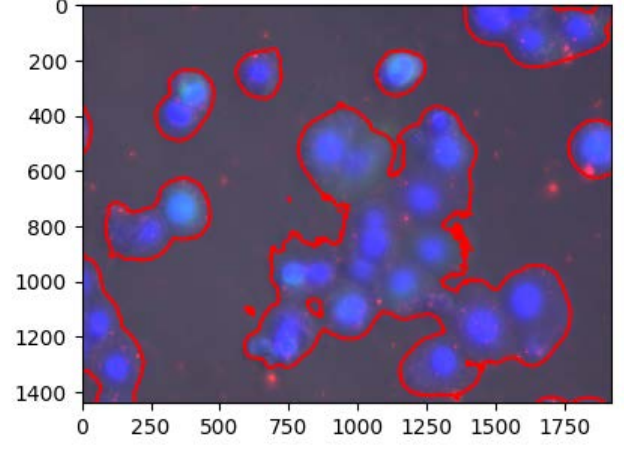}
         \caption{Raw image with mask overlay.}
         \label{fig:contour_original_mask}
     \end{subfigure}
     \begin{subfigure}[b]{0.3\textwidth}
         \centering
         \includegraphics[width=\textwidth]{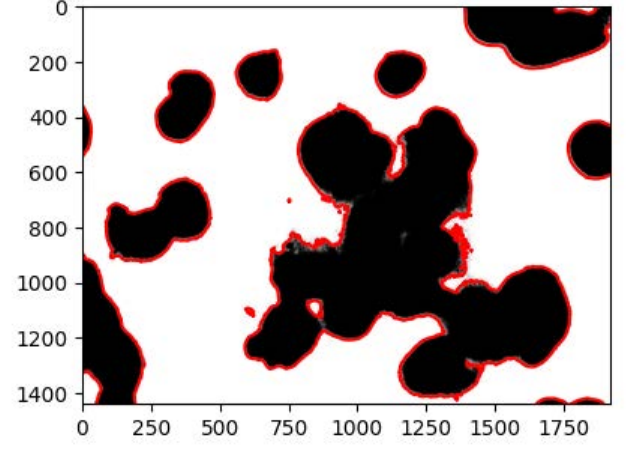}
         \caption{Mask with labeled boundaries.}
         \label{fig:mask_contour}
     \end{subfigure}
     \begin{subfigure}[b]{0.3\textwidth}
         \centering
         \includegraphics[width=\textwidth]{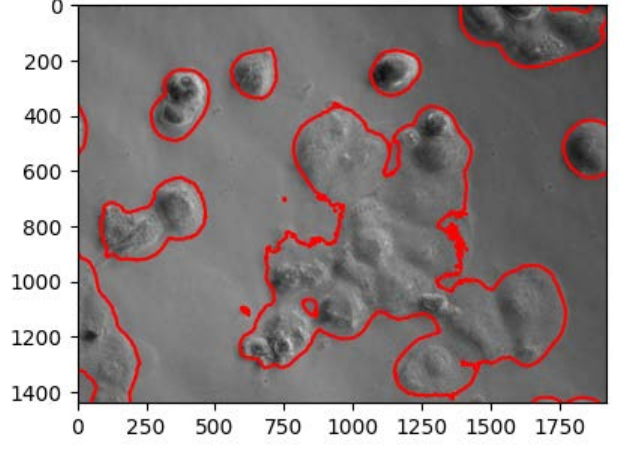}
         \caption{Phase image with mask overlay.}
         \label{fig:phase_contour_mask}
     \end{subfigure}
        \caption{Fluorescent images of siRNA (red), nuclei (blue), and endosomes (green). We very the mask accuracy using a red line to denote the estimated locations of the cell membranes. (a) The original fluorescent image captured at time t = 4 hours with mask overlay to verify the cells are contained within the mask. (b) User defined mask using thresholding technique in Section \ref{sec:imageanalysis} with boundary labeled. (c) Phase image of original fluorescent image with mask overlay to verify that cells are contained inside the mask.}
        \label{fig:mask_overlays}
\end{figure*}

\section{Methods}
\label{sec:methods}

\subsection{Image Analysis}
\label{sec:imageanalysis}

In this paper, we implemented a computational image analysis approach to quantify the endosomal escape of siRNA using fusogenic peptides. This method involves the processing of high-resolution TIF files of fluorescent microscopy images obtained during experiments detailed in Section \ref{sec:experimental_methods}. The cell images contain color channels: red (corresponding to siRNA), green (corresponding to endosomes), and blue (corresponding to cell nuclei). 

Each petri dish contains multiple cells. To identify the general locations of the individual cells, a user-defined threshold is applied to create a binary mask to capture the blue channel (nuclei) in the image in Figure~\ref{fig:mask}. The binary mask is generated where the intensities of the blue labeled objects exceed the threshold. This mask overlaps with the cellular structure present in the fluorescent microscopy images in Figure \ref{fig:contour_original_mask}. The mask boundaries are outlined in red in Figure~\ref{fig:mask_contour} which is used to segment the original image in Figure~\ref{fig:remove_sirna}. The interactive tool, \textit{ipywidgets} \cite{mease2018bringing}, is then used to allow fine-tuning of the threshold and Gaussian smoothing parameters. Users can visually adjust the quality of the mask to ensure that it accurately represents the desired regions. The generated mask is used to filter the cellular region of the image. Only labeled objects corresponding to the inside of the regions defined by the mask are retained, while other regions are set to zero, which implies the removal of floating siRNA as shown in Figure~\ref{fig:remove_sirna}. Additionally, we overlay the mask on the original image to verify that we are masking the petri dish's regions containing cells and their perceived cell membrane. We verify the mask accuracy using the phase image in Figure \ref{fig:phase_contour_mask}.

The resulting masked image contains the cells, their nuclei, and siRNA only within the cell membrane, i.e. removing all floating siRNA. We use the masked images to quantify endosomal escape over a period of 4 hours and treat these values as our experimental data. Using OTSU's method \cite{Otsu1979ATS}, we differentiate between each color channel by creating binary masks to count the red, green, and blue labeled objects. These masks enable the identification of areas where only red is present in the cell membrane or where red and green overlap. The latter case implies that siRNA are contained in the endosome.

In Figure~\ref{fig:raw_image}, the \textit{blue} color denotes nuclei, the \textit{red} color outside of the \textit{green} color denotes endosomal escape of siRNA, \textit{green} denotes the endosome, and overlapping \textit{red} and \textit{green} denotes the co-localization of siRNA and endosome (i.e. endosomal escape has not occurred). The primary outcome measure is the percentage of each color representing endosomal escape, computed by dividing the number of red pixels outside endosome (green pixels) by the total number of red pixels within the cells. This metric provides a quantitative measure of endosomal escape based on the pixel counts of fluorescent microscopy images.

\subsection{Data Analysis}
\label{dataanalysis}

\begin{figure}[h]
    \centering
    \includegraphics[width=0.75\linewidth]{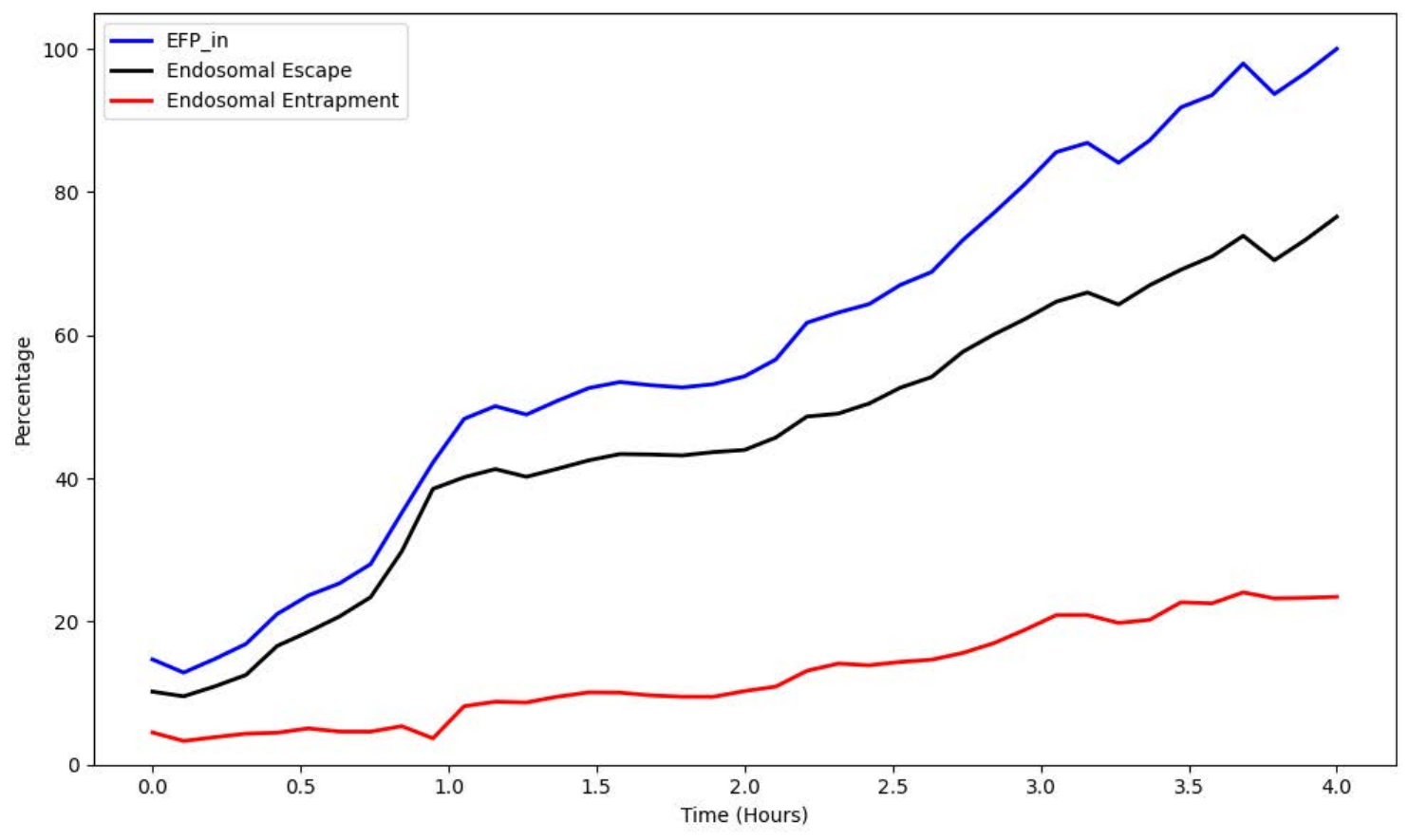}
    \caption{Experimental data extracted using the fully automated model. Each curve represents the components described in Section \ref{dataanalysis}. The \textit{blue} curve denotes EFP$_{\text{in}}$(siRNA and fusogenic peptides encapsulated inside the endosome). The \textit{black} curve denotes EE (Endosomal Escape). The \textit{red} curve denotes endosomal entrapment.}
    \label{fig:data-to-use}
\end{figure}

We implement the method proposed in Section \ref{sec:imageanalysis} to compute the percentage of endosomal escape in a series of 40 frames, where one frame is captured every 5 minutes. Figure~\ref{fig:data-to-use} shows the extracted experimental data describing the endosomal escape percentage in each image from time $t=0$ to $t=4$ hours. We define cellular uptake (EFP$_{\text{in}}$), siRNA entering the cell, as the total number of red objects (siRNA) at time $t=0$. Over time, this is denoted by the \textit{blue} curve. Endosomal escape (EE) is denoted by the \textit{black} curve. Degradation is represented by the difference between EFP$_{\text{in}}$ and EE, indicating the amount of siRNA that is unable to escape from the endosome at a given time t, as shown by the \textit{red} curve. The time series data indicates a positive correlation between the volume ratio of peptides used and the percentage of endosomal escape, suggesting that prolonged exposure to peptides enhances the efficiency of siRNA delivery.

Endosomal escape is an inherently random biological process. Following cellular uptake, endosomes move dynamically within the cell, internalizing nearby particles, including our peptide-siRNA complexes. However, it remains unclear how long each endosome takes to internalize all incoming particles or which endosomes will encapsulate specific peptide-siRNA complexes. This randomness influences the process of endosomal escape, which depends on the duration of particles remaining within the endosome. Consequently, siRNA release occurs unpredictably across cells and over varying time intervals.

To account for this inherent randomness, we employ stochastic compartmental models, which capture the variability in endosomal escape across the numerous endosomes present in the many cells within the petri dish. The use of stochastic simulation algorithms is particularly advantageous, as they model the random interactions between compartments and their effects on the changes in compartment populations.

\subsection{Simulation Methods}
\label{SSA}

The stochastic simulation algorithm (SSA) is traditionally used to model chemical reaction systems in cellular and molecular biology \cite{Qian2012,Gillespie1976, Gillespie1977,Hattne2005, Erban2009, Alberts2007,Andrews2004}. More specifically, these systems model the time evolution of populations of molecules with respect to their interactions at infinitesimal time intervals. In this work, we use the SSA framework developed by Daniel T. Gillespie in the 1970s \cite{gillespie1977exact} to model the percentage of endosomal escape over a 4-hour time period. The SSA is particularly useful for this system because of the discrete behavior of the individual intracellular organelles and particles. The SSA allows us to take into account the intrinsic unpredictability of these interactions. The algorithm simulates each reaction event on a probabilistic basis by first determining when the next reaction will occur (exponentially distributed) and then deciding which reaction will take place. These are determined by the relative propensity of each reaction in our system, i.e. the propensity function represents the likelihood of each reaction occurring at a given moment, given the current state of the system. 

For this work, we treat the pixels as molecules and the reactions to be compartment-location-based transitions of the pixels over time. Let the compartments be defined as FP (fusogenic peptides), siRNA, EFP$_{\text{in}}$ (siRNA and fusogenic peptides encapsulated inside the endosome), EFP$_{\text{trap}}$ (endosome entrapment), and EE (endosomal escape). Note that the siRNA compartment represents the initial state of the small interfering RNA molecules involved in the reaction, EFP$_{\text{in}}$ represents the early stage of uptake of siRNA by endosome using the fusogenic peptides, EE represents the siRNA that escape the endosome, and EFP$_{\text{trap}}$ represents the amount of siRNA trapped inside the endosome and eventually degrades. The reaction system is defined by the following reactions:

\begin{align}
    \text{FP} + \text{siRNA} &\xrightarrow{\mu} \text{EFP}_{in}  \label{eq:1}\\
\text{siRNA} &\xrightarrow{\omega} \emptyset  \label{eq:2} \\
\text{EFP}_{\text{in}} &\xrightarrow{\sigma} \text{EE}  \label{eq:3}\\
\text{EFP}_{in}&\xrightarrow{d_{exit}} \text{EFP}_{\text{trap}}  \label{eq:4} 
\end{align}

It is important to understand that reaction \eqref{eq:1} represents the interaction of fusogenic peptides  with siRNA and ${\mu}$ is the rate at which the combination is taken up by the endosome. Reaction \eqref{eq:2} refers to the scenario when no fusogenic peptides are present, hence the uptake is bare minimum. This reaction relies on $\omega$, which is set to zero in our simulations because we know that peptides are present in our experiments. This reaction can be used to generate a base simulation. Reaction \eqref{eq:3} models the release of siRNA from EFP$_{\text{in}}$, denoted by EE, where $\sigma$ is the rate of release. Reaction \eqref{eq:4} models the rate $d_{exit}$ at which EFP$_{\text{in}}$ degrades.

Experimentally, as detailed in Section \ref{sec:experimental_methods}, the ovarian cancer cells are incubated with a ratio of 80:1 fusogenic peptides and siRNA; i.e. each well in a 96 well plate contains concentrations of 1.2 $\mu L$ of siRNA and 8.27 $\mu L$ of fusogenic peptides (FP). Equation (\ref{ratio}) denotes the ratio of peptides to siRNA, 
\begin{equation}\label{ratio}
    \text{Ratio} = \frac{\text{FP}}{\text{siRNA}},
\end{equation}
hence the numerical ratio used in the model for the concentration of fusogenic peptides to siRNA is 6.891.

Let the initial conditions for each compartment be defined by the vector in Table 1,
\begin{table}[htbp]
\begin{center}
\resizebox{1\textwidth}{!}{
\begin{tabular}{ccccc}
\hline
\textbf{FP} & \textbf{siRNA} & \textbf{EFP$_{\text{in}}$} & \textbf{EE} & \textbf{EFP$_{\text{trap}}$}  \\
\hline
$\text{siRNA} \times \text{Ratio}$ & $120$  & $\text{EFP}_{\text{in}} \times \text{siRNA}$ & $\text{EE} \times \text{siRNA}$ & $(\text{EFP}_{\text{in}} - \text{EE}) \times \text{siRNA}$ \\
\hline
\end{tabular}
}
\end{center}
\caption{Initial conditions for each compartment.}
\label{tab:initial_conditions}
\end{table}
where Ratio in Equation (\ref{ratio}). Since EFP$_{\text{in}}$ and $EE$ denote the cellular uptake and endosomal escape percentages, we let the initial conditions be determined by the experimental data at time $t=0$. Let the initial volume of siRNA be 120 $\mu L$; thus each initial compartment amount will be multiplied by $120$. Let the initial condition for $FP = 120 \times$ Ratio. In this work, we assume that the rate parameter $\omega$ for reaction \eqref{eq:2} is $0$. That is, siRNA does not degrade on its own since they are within a peptide-siRNA complex. In Section \ref{inference}, we detail the process for inferring the distribution of the rate parameters $\mu$, $\sigma$ and $d_{exit}$ in reactions \eqref{eq:1}, \eqref{eq:3} and \eqref{eq:4}, respectively. 

\section{Inference Methods and Simulations}
\label{inference}

\subsection{Parameter Inference}
\label{estimate_params}

The reaction rates of a biochemical reaction network are of utmost importance. In this work, the experimental data shows that the population of each compartment is not constant over time. Due to the natural random release of siRNA into the cytosol of each cell, we assume that the rates fluctuate at each time point. To account for variable reaction rates over time, we must estimate the rate of change with respect to the associated propensity function that is computed at each time step in the SSA. In \cite{Warne2019}, the authors describe estimating varying reaction rates using Bayesian computation, where they solve the inverse problem of choosing the best-fit reaction rates for the system.

Let $\textbf{X}_{obs} = [\textbf{X}(t_1),\textbf{X}(t_2),\dots,\textbf{X}(t_n)]$ be the experimental observations at time points $t_1,\dots t_n$, where $n=39$. (Recall that the fluorescent images are captured every $\Delta = 5$ minutes from time $t=0$ to $t=4$ hours.) $\textbf{X}$ denotes the vector containing components
\begin{equation}
    \frac{\text{EFP}_{\text{in}}(t_n)}{dt} = \frac{\text{EFP}_{\text{in}}(t_{n+1})-\text{EFP}_{\text{in}}(t_n)}{\Delta},
\end{equation}
\begin{equation}
    \frac{\text{EE}(t_n)}{dt} = \frac{\text{EE}(t_{n+1})-\text{EE}(t_n)}{\Delta}
\end{equation}
and 
\begin{equation}
   \frac{ \text{EFP}_{\text{trap}}(t_n)}{dt} = \text{EFP}_{\text{in}}(t_n) - \text{EE}(t_n).
\end{equation}
That is, EFP$_{\text{in}}(t_n)$, EE$(t_n)$ and $EFP_{\text{trap}}(t_n)$ are rates of change for each time step of the experimental data. 
 
PyMC \cite{patil2010pymc} is a Python library for probabilistic programming that enables users to construct and analyze Bayesian statistical models. It provides a suite of modern Bayesian tools for model specification, fitting, and diagnostics. PyMC implements advanced Markov Chain Monte Carlo (MCMC) algorithms for sampling from a posterior distributions.

We use a hierarchical Bayesian approach that estimates the reaction rate for the selected reaction at each time step. This approach builds a statistical model of the experimental endosomal escape data by splitting the inference into steps based on conditional distributions. First, we define the global hyperpriors ($\mu^*,\sigma^*,$ and $d_{exit}^*$) for each of the reaction rates, representing the overall trend for each rate over all time steps. 

Let the global hyperpriors for $\mu$ and $\sigma$ be Cauchy distributed. Using the experimental data for EFP$_{\text{in}}$ and EE, we estimate their respective distributions using their differences over time described in Equations 6 and 7. We fit each set of differences to a Cauchy distribution using \textit{fitdistr} in the R package \textit{MASS} used for fitting probability distributions.
Hence, $\mu^* \sim Cauchy(0.326,0.219)$ and $\sigma^* \sim Cauchy(0.238,0.137)$. Let the global hyperpriors for $d_{exit}$ be normally distributed. Using the experimental data in Equation 8, $d_{exit}^* \sim Normal(0.036,0.1)$. Then we define the $n$ local priors for each time step to be normally distributed with a small standard deviation ($\tau_\mu,\tau_\sigma,\tau_{d_{exit}}$) = ($0.1,0.1,0.1$) around the sampled global hyperpriors; $\mu(t)\sim N(\mu^*,\tau_\mu)$, $\sigma(t)\sim N(\sigma^*,\tau_\sigma)$, and $d_{exit}(t)\sim N(d_{exit}^*,\tau_{d_{exit}})$. The likelihood functions for each parameter are defined as
\begin{equation}
    p(\text{EFP}_{\text{in}}|\mu) = \prod^n_{t=1}N(\text{EFP}_{\text{in}}(t)|\mu(t),\gamma_\mu),
\end{equation}
\begin{equation}
    p(\text{EE}|\sigma) = \prod^n_{t=1}N(\text{EE}(t)|\sigma(t),\gamma_\sigma),
\end{equation}
\begin{equation}
    p(\text{EFP}_{\text{trap}}|d_{exit}) = \prod^n_{t=1}N(\text{EFP}_{\text{trap}}(t)|d_{exit}(t),\gamma_{d_{exit}}),
\end{equation}
where $(\gamma_\mu,\gamma_\sigma,\gamma_{d_{exit}})$ = ($0.3,0.3,0.3$) are small standard deviations about the mean.
Thus the joint likelihood function is then defined as

\begin{eqnarray}
 p(data|\mu,\sigma,d_{exit}) &=& \prod^{n}_{t=1}N(\text{EFP}_{\text{in}}(t)|\mu(t),\gamma_\mu) \nonumber\\ 
      & & N(\text{EE}(t)|\sigma(t),\gamma_\sigma) N(\text{EFP}_{\text{trap}}(t)|d_{exit}(t),\gamma_{d_{exit}}).
\end{eqnarray}

PyMC then uses a Markov Chain Monte Carlo (MCMC) sampling algorithm to estimate the posterior distributions of all the parameters. We implement 6000 samples with an additional 1000 to tuning iterations to adjust the sampling algorithm and run 5 independent sampling chains to validate the posterior estimates. Algorithm \ref{alg:pyMC} lists the steps in a concise manner. 

\begin{algorithm}
\caption{PyMC Posterior Model}\label{alg:pyMC}
\begin{algorithmic}
\State n = number of time steps
\Ensure Global Hyperparameters
\State $\mu^* \sim Cauchy(\alpha_\mu,\beta_\mu)$
\State $\sigma^* \sim Cauchy(\alpha_\sigma,\beta_\sigma)$
\State $d_{exit}^* \sim Normal(\alpha_{d_{exit}},\beta_{d_{exit}})$
\Ensure Sample local Priors for each time step $t_n$
\State $\mu(t)\sim N(\mu^*,\tau_\mu)$
\State $\sigma(t)\sim N(\sigma^*,\tau_\sigma)$
\State $d_{exit}(t) \sim N(d_{exit}^*,\tau_{d_{exit}})$

\Ensure Define joint likelihood

\State $p(data|\mu,\sigma,d_{exit}) = \prod^{n}_{t=1} p(\text{EFP}_{\text{in}}|\mu)\cdot p(\text{EE}|\sigma)\cdot p(\text{EFP}_{\text{trap}}|d_{exit})$
\Require Draw samples from the posterior $P(\mu,\sigma,d_{exit}|data)$

\State Posterior Means at each time step: 
    \State $\hat{\mu}(t) = \mathbb{E}[\mu(t)], \hat{\sigma}(t)=\mathbb{E}[\sigma(t)], \hat{d}_{exit}(t)=\mathbb{E}[d_{exit}(t)] $

\end{algorithmic}
\end{algorithm}

We compute the posterior distributions for each of the parameters $\mu$, $\sigma$, and $d_{exit}$ at each time step $t$ and let the posterior mean at each time step be the estimate for the rate. To generate sample paths using the SSA, we use the estimated rate at each time step of the associated reaction. The posterior distributions for the inferred parameters at each time step are shown in Figure~\ref{fig:appendix:pymc} in the Appendix. 

We observed significant variations in parameter rates across different peptide ratios in our experimental data. To ensure the model remains broadly applicable to a wide range of peptide-siRNA ratios, we adopted a more generalized approach for parameter selection rather than assuming constant rates. This flexibility allows the model to adapt to varying experimental conditions without being constrained by fixed rate assumptions. Biologically, the reaction rates cannot be constant because one petri dish contains many cells. We use time-dependent rates because endosomal membrane ruptures and destabilization depend on the quantity of peptides present, and these disruptions are not uniform over time. As a result, the release of siRNA varies at each time step in real biological systems. Hence, variable reaction rates are most representative of endosomal escape over time. 

\subsection{Simulating Realizations of the SSA using Inferred Rate Parameter \\ Estimates \label{SSA_estimate}}

\begin{figure}[htbp]
    \centering
    \includegraphics[width=1\linewidth]{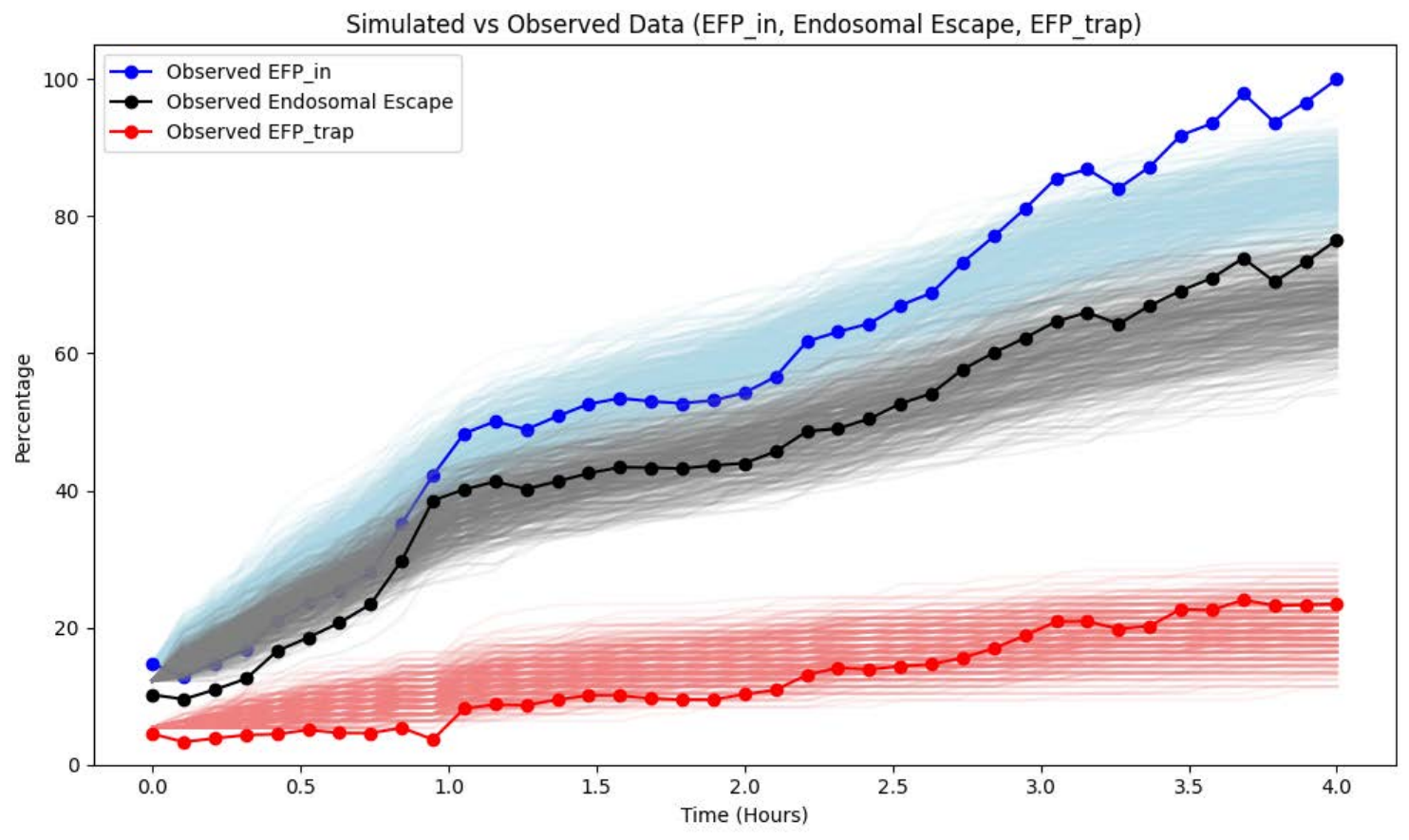}
    \caption{The observed experimental data (bold blue, black, and red lines labeled in the plot legend) plotted with the corresponding simulations (the light color blue, black, and red curves).
    Each curve represents the components described in Section \ref{dataanalysis}. The \textit{blue} and \textit{light blue} curves denote EFP$_{\text{in}}$. The \textit{black} and \textit{gray} curves denote EE. The \textit{red} and \textit{light red} curves denote endosomal entrapment.}
    \label{fig:ssa_sims}
\end{figure}

We use the posterior means for each time step as the reaction rates $\mu$, $\sigma$, and $d_{exit}$, along with the initial conditions in Section \ref{SSA} to simulate realizations of each component using the SSA \cite{Gillespie1976}. Figure \ref{fig:ssa_sims} displays 6000 realizations of the SSA for EFP$_{\text{in}}$ (\textit{light blue} curves), EE (\textit{gray} curves), and EFP$_{\text{trap}}$ (\textit{light red} curves) compared to the experimental data for each component, \textit{blue}, \textit{black}, and \textit{red}, respectively.

\section{Sensitivity Analysis}
\label{SA}

To understand how variations in model parameters influence the outcomes in different system compartments, we performed sensitivity analysis using two widely recognized methods: the Fourier Amplitude Sensitivity Test (FAST)\cite{christopher2002identification} and Sobol sensitivity analysis \cite{zhang2015sobol}. These approaches help identify the key parameters driving model behavior and provide insights into their relative contributions. FAST is a global sensitivity analysis that estimates how much each input variable contributes to the variance of the model output using Fourier transforms. It provides two key sensitivity indices: 
\begin{itemize}
    \item \textbf{First-order indices (S1):} These represent the direct effect of a single parameter on the output, independent of interactions with other parameters.
    \item \textbf{Total sensitivity indices (ST):} These capture the combined effect of a parameter, including its interactions with all other parameters.
\end{itemize}
FAST is computationally efficient and suitable for models with many inputs. Sobol decomposes the total variation in the output of the model to quantify the contribution of each input variable and their interactions. It provides:
\begin{itemize}
    \item \textbf{First-order indices (S1):} Similar to FAST, these measure the contribution of a single parameter to the variance of the output without considering interactions.
    \item \textbf{Total-order indices (ST):} These encompass both the direct effects of a parameter and its interactions with others, offering a more comprehensive picture of parameter influence.
\end{itemize}
Unlike FAST, Sobol explicitly quantifies the contribution of higher-order interactions. However, it is computationally intensive due to its reliance on extensive Monte Carlo sampling.\\

\begin{figure*}
     \centering
     \begin{subfigure}[b]{0.45\textwidth}
         \centering
         \includegraphics[width=1\textwidth]{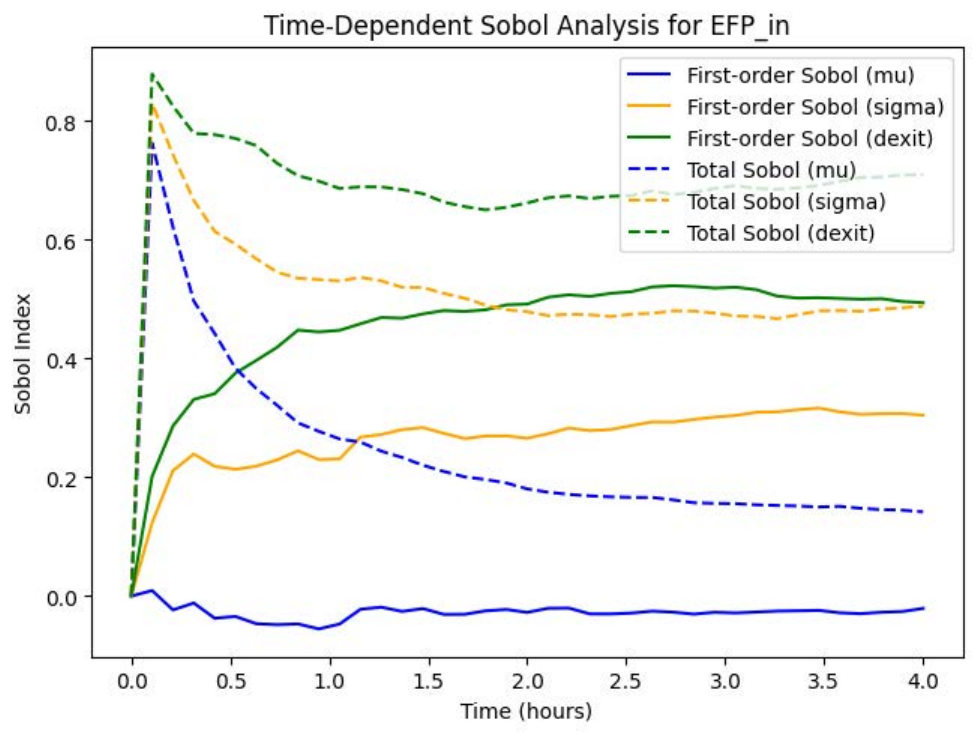}
         \caption{Time-Dependent Sobol Sensitivity Analysis for EFP$_{\text{in}}$}
         \label{fig:sens_efp_in}
     \end{subfigure}
     \begin{subfigure}[b]{0.45\textwidth}
         \centering
         \includegraphics[width=1\textwidth]{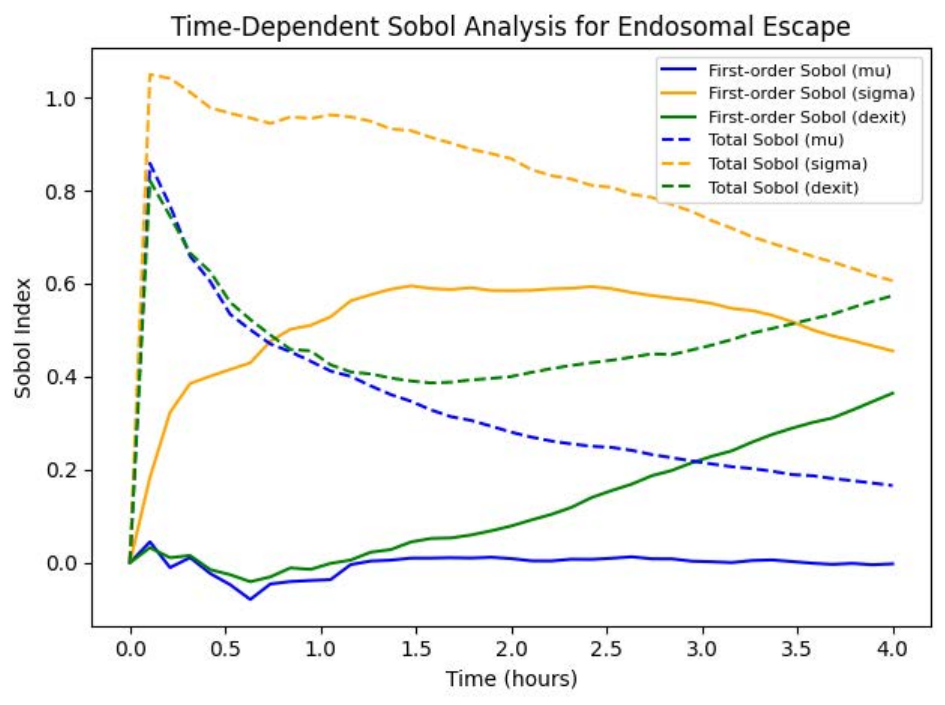}
         \caption{Time-Dependent Sobol Sensitivity Analysis for EE}
         \label{fig:sens_ee}
     \end{subfigure}
     \\
     \begin{subfigure}[b]{0.45\textwidth}
         \centering
         \includegraphics[width=1\textwidth]{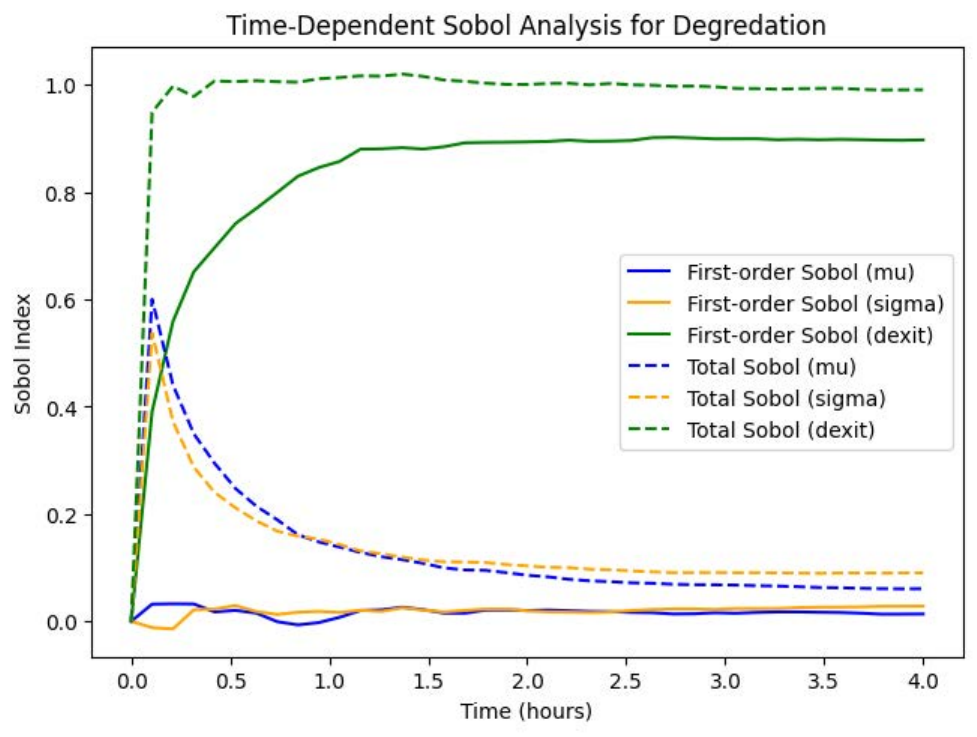}
         \caption{Time-Dependent Sobol Sensitivity Analysis for $d_{exit}$.}
         \label{fig:sens_deg}
     \end{subfigure}
        \caption{Time-Dependent Sobol sensitivity analysis for each of the compartments EFP$_{\text{in}}$, EE, and EFP$_{\text{trap}}$ detailed in Section \ref{TimeSA}. The \textit{blue} curves represent the affects of parameter $\mu$, the \textit{yellow} curves represent the affects of parameter $\sigma$, and the \textit{green} curves represent the affects of parameter $d_{exit}$. The dashed lines denote  Total order Sobol and the solid lines denote the First-order Sobol.  }
        \label{fig:sensitivity}
\end{figure*}

\begin{figure*}
     \centering
     \begin{subfigure}[b]{0.45\textwidth}
         \centering
         \includegraphics[width=\textwidth]{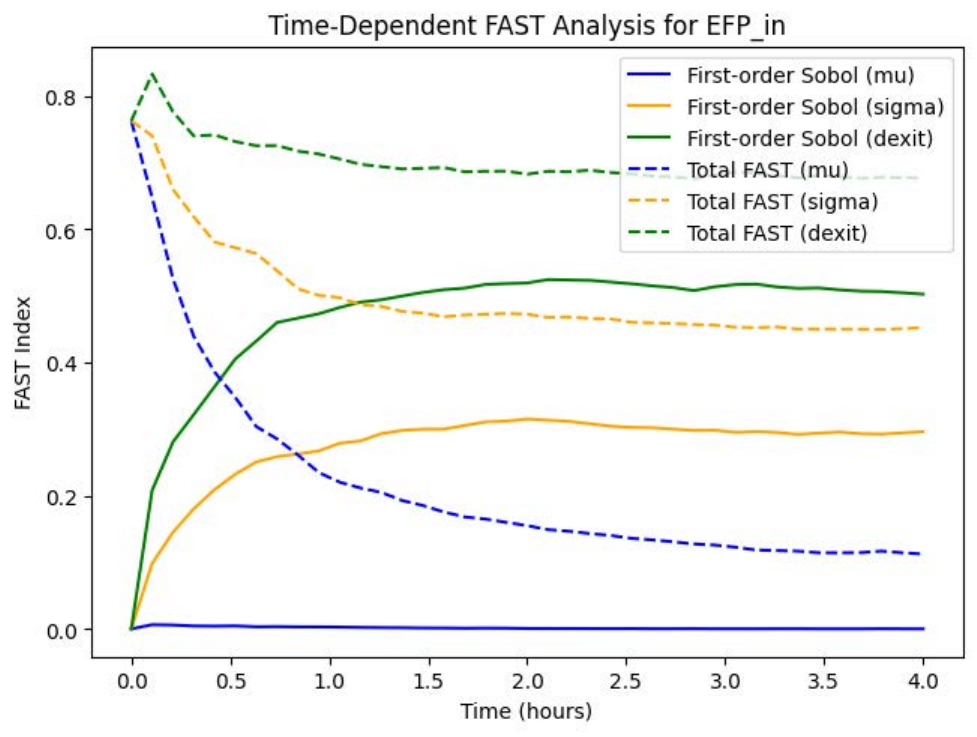}
         \caption{Time-Dependent FAST Sensitivity Analysis for $EFP_{in}$}
         \label{fig:fast_sens_efp_in}
     \end{subfigure}
     \begin{subfigure}[b]{0.45\textwidth}
         \centering
         \includegraphics[width=\textwidth]{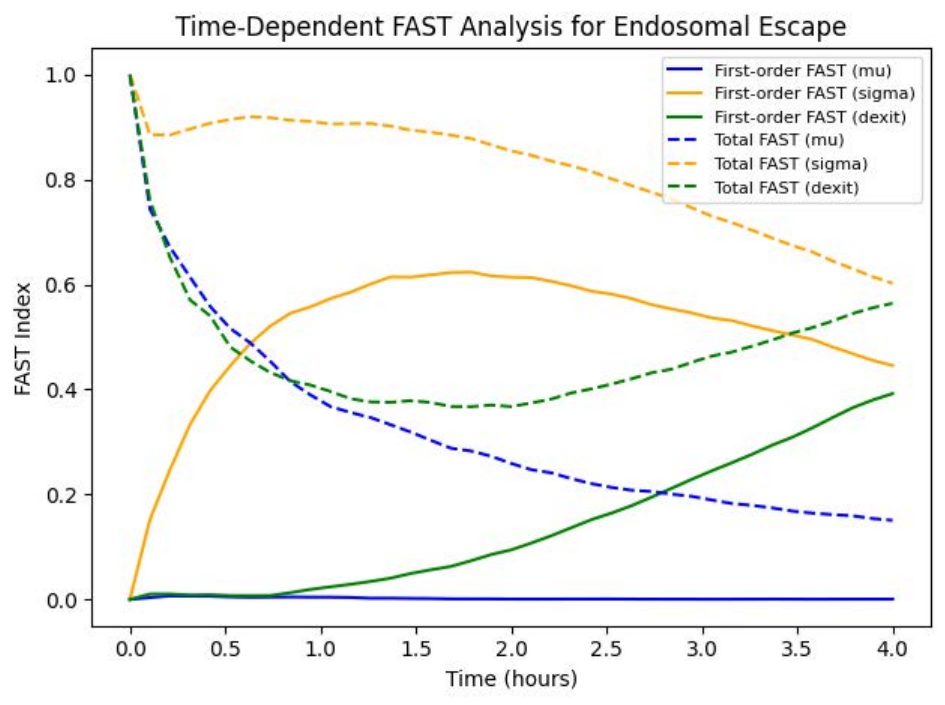}
         \caption{Time-Dependent FAST Sensitivity Analysis for $EE$}
         \label{fig:fast_sens_ee}
     \end{subfigure}
     \\
     \begin{subfigure}[b]{0.45\textwidth}
         \centering
         \includegraphics[width=\textwidth]{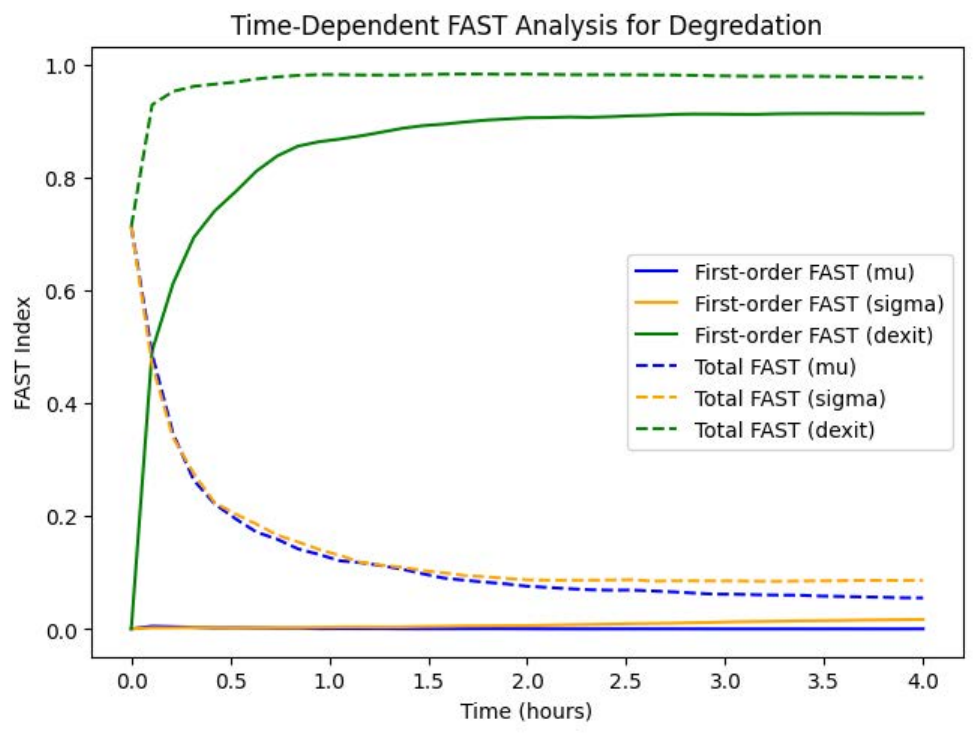}
         \caption{Time-Dependent FAST Sensitivity Analysis for $d_{exit}$.}
         \label{fig:fast_sens_deg}
     \end{subfigure}
        \caption{Time-Dependent FAST sensitivity analysis for each of the compartments $EFP_{in}$, $EE$, and $EFP_{trap}$ detailed in Section \ref{TimeSA}. The \textit{blue} curves represent the affects of parameter $\mu$, the \textit{yellow} curves represent the affects of parameter $\sigma$, and the \textit{green} curves represent the affects of parameter $d_{exit}$. The dashed lines denote  Total FAST and the solid lines denote the First-order FAST.  }
        \label{fig:FAST_sensitivity}
\end{figure*}

We study the affect of parameters  ($\mu$, $\sigma$, $d_{exit}$) on the output of model using the FAST and Sobol methods on the simulated data, with a particular emphasis on the predicted endosomal escape percentage. The results of the sensitivity analysis methods are presented in Figure \ref{fig:sensitivity} and \ref{fig:FAST_sensitivity}. 

\subsection{Time-Dependent Sensitivity Analysis}
\label{TimeSA}

The time-dependent sensitivity analysis provides valuable insights into the relative importance of the model parameters and how their influence evolves over time. Such an analysis is crucial for guiding experimental design and model refinement, particularly in optimizing parameters to improve the accuracy and robustness of simulations for endosomal escape. Both methods confirm the dominance of \( d_{\text{exit}} \) in degradation and \( \sigma \) in endosomal escape, highlighting these as critical parameters for optimization in cellular system modeling.\\

Figure \ref{fig:sensitivity} and \ref{fig:FAST_sensitivity} illustrate the results of Sobol and FAST sensitivity analysis, respectively, applied to the time evolution of the system’s behavior. Each focuses on the dynamics of EFP$_{\text{in}}$, EE, and EFP$_{\text{trap}}$ over a simulation period of 4 hours. This analysis was performed to identify the influence of three key model parameters $\mu$, $\sigma$, and $d_{exit}$ on the variance of the system's output. The Sobol and FAST index values are plotted over time $t=0$ to $t=4$ hours. It is important to note that the Sobol and FAST sensitivity indices fall within the range of 0 to 1, where values near 0 indicate minimal influence on the system's output and values near 1 signify a significant influence on the system's output. If any of the curves surpasses 1 it means it is affected by interactions with other parameters. These indices quantify the contributions that each parameter has on the variance of the output. First-order Sobol/FAST indices (S1) measure the independent contribution of a parameter. Total Sobol/FAST indices (ST) account for the combined effects of a parameter, including interactions with other parameters. Parameters $\mu, \sigma$ and $d_{exit}$ are distinguished using separate colors, \textit{blue}, \textit{orange}, and \textit{green}, respectively. The solid lines represent first-order indices (S1) and the dashed lines represent total indices (ST).\\

We observed that the Sobol and FAST sensitivity analysis graphs exhibit very similar patterns, which strongly supports the conclusion of a significant relationship between the parameters and the output. For EFP$_{\text{in}}$ in Figures~\ref{fig:sens_efp_in} and \ref{fig:fast_sens_efp_in}, the parameter $d_{exit}$ is the dominant contributor during the entire 4 hours time period, with a total Sobol index ($ST$) between 0.6 and 0.8, highlighting its critical role in the system's EFP$_{\text{in}}$ dynamics as a whole.\\

In contrast, for \textit{Endosomal Escape} in Figures~\ref{fig:sens_ee} and \ref{fig:fast_sens_ee}, the parameter $\sigma$ is the dominant contributor during the entire 4 hours time period, with a total Sobol index ($ST$) between 1 and 0.6, highlighting its critical role in the system's \textit{Endosomal Escape} dynamics as a whole. Interestingly, both the First-order and Total Sobol indices for $\mu$ and $d_{exit}$ are about the same until after hour 1.5 suggests that these parameters contribute primarily as independent (non-interacting) effects on the model output within this time frame. For \textit{Degradation} in Figures~\ref{fig:sens_deg} and \ref{fig:fast_sens_deg}, the parameter $d_{exit}$ exhibits consistently high sensitivity across all time phases, particularly dominating the early and late phases, with $\mu$ and $\sigma$ having moderate influence in the initial 0.5 hour. \\

The parameter $\sigma$, while crucial for \textit{Endosomal Escape}, has a lesser impact on EFP$_{\text{in}}$ and even lesser impact on \textit{Degradation}, emphasizing the distinct roles these parameters play in different outputs. This comparison underscores the importance of $\sigma$ in the first 2 hours, while $d_{exit}$ drives the key dynamics of EFP$_{\text{in}}$ and \textit{Degradation}. These insights provide valuable guidance for parameter tuning and optimization in cellular system modeling.

\section{Discussion}
\label{Discussion}

Endosomal escape remains a major barrier to nanotechnology delivery of nucleic acid cargo, and therefore, it is extensively studied in research. This mathematical framework can be used to support these investigations and quantify the endosomal escape of siRNA through peptide-based delivery.  Current methods for analyzing endosomal escape typically rely on qualitative analysis of fluorescent images and end-point assays. We build off these limitations to design a reproducible model that tracks endosomal escape dynamics and predicts the endosomal escape efficiency of peptide-based carriers, streamlining the design and experimental process. Therefore, this model will provide a better indication of the escape capability of carriers \textit{in vitro}, which will improve the translation of these nanotechnologies into \textit{in vivo} studies and clinically, helping to further the gene therapy delivery field.  

In this work, we developed a comprehensive framework that 1) utilizes image processing techniques to automate the extraction of object data related to endosomal escape from time-series fluorescent microscopy images, 2) estimates the rate of endosomal escape over time using Bayesian approaches, and 3) simulates synthetic data that replicates the experimentally collected endosomal escape data. Note that previous computational endosomal escape methods relied on only the fluorescence intensity of the labeled particle over time. Our method improves on this methodology by incorporating experimental factors into the model and instead counting the fluorescent pixels that are associated with each particle. 

This model allows users to generate synthetic data by inputting experimental parameters. By validating the synthetic data against experimental results, researchers can ensure the model accurately reflects observed behavior. Once validated, the model enables experimentalists to predict endosomal escape outcomes under different conditions without conducting additional experiments. For example, users can explore how the percentage of endosomal escape changes over extended periods (e.g., 8 or 10 hours) or evaluate the effects of altering initial siRNA and peptide ratios. By simply adjusting the simulation parameters, the model provides insights into how varying experimental conditions influence endosomal escape, reducing the need for additional costly and time-consuming experiments.

\subsection{Model Limitations}
\label{limits}

The Bayesian inference component of our framework is adaptable, allowing users to adjust it based on their collected experimental data. Because the model employs a hierarchical structure, the global hyperpriors must be fitted to the experimental data; however, this process needs to be performed only once. After validation, rate estimates are drawn from the posterior distributions, which naturally account for fluctuations in rate parameters. These fluctuations are further refined during the error optimization step, ensuring that the simulations closely align with the experimental data.

\subsection{Implications for Future Work}
\label{futurework}

When dealing with multiple experiments, it is beneficial to apply Bayesian methods to all time-series datasets, estimating a unique set of parameters for each experimental condition. This approach enhances the model’s robustness, enabling it to accommodate varying siRNA-to-peptide ratios effectively. Incorporating machine learning methods can further extend the model’s capabilities by predicting endosomal escape percentages over time under different experimental conditions. While this adds an additional layer of complexity, it provides valuable insights for experimentalists, enabling more precise predictions and reducing the need for additional experiments.

There are a myriad of different image processing tools that can be employed to analyze fluorescent microscopy images of endosomal escape data. In this work, we utilized binary masks and Gaussian filters, but these methods can be updated or expanded upon based on the latest advancements in the literature.

For instance, other types of masks such as adaptive thresholding masks, edge-detection masks (e.g., Sobel or Canny), or region-growing masks can be applied to segment and analyze specific regions of interest within the images. These masks can help isolate distinct features like endosomes or fluorescently labeled siRNA more accurately, depending on the complexity and variability of the data.

Additionally, various image filters can be employed to enhance specific aspects of the microscopy images. Median filters are effective for reducing noise while preserving edges, whereas Laplacian filters can enhance edge detection to better delineate boundaries. High-pass filters can amplify small details, and band-pass filters can isolate features within a specific size range, allowing researchers to focus on relevant structures like endosomal vesicles. By tailoring the choice of masks and filters to the experimental requirements, the image processing pipeline can be fine-tuned to capture the critical elements of endosomal escape with greater precision.

The data for this work can be accessed at DOI
10.5281/zenodo.15288765, and the code is available at \href{https://github.com/Nisha1803/Methodology_Endosomal_Escape}{GitHub}.

\section{Acknowledgments}

This research was supported by the Clemson University CU-Fellows faculty research support grant and 
National Science Foundation Faculty Early Career Award Development Program Under NSF Award \#204669.\\

\appendix
\section{Posterior Estimates for Reaction Rates: PyMC}
\label{appendix}

\begin{figure*}[h]
    \centering
    \includegraphics[width=1\linewidth]{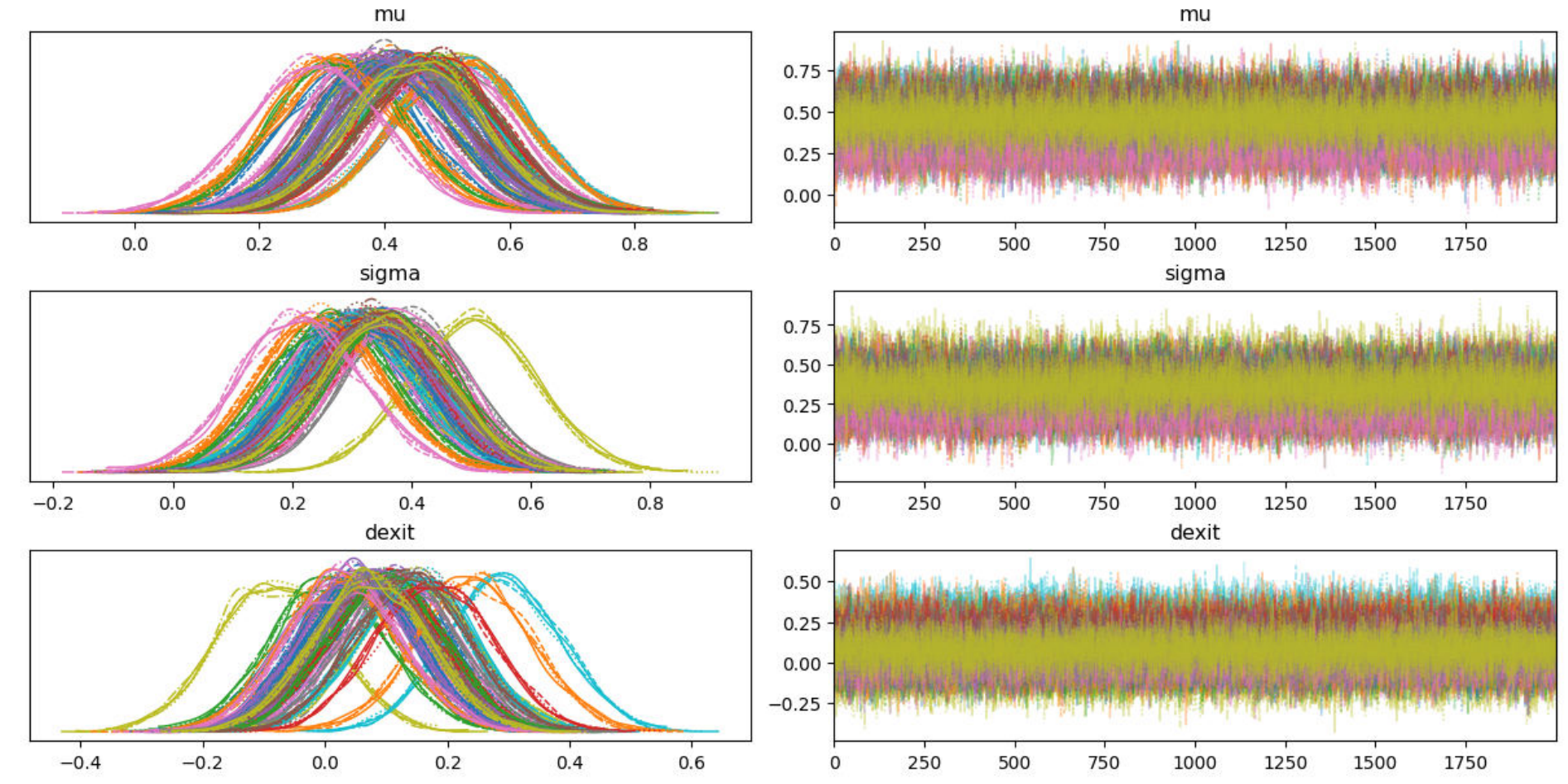}
    \caption{ PyMC posterior parameters for each variable over each time step. Each color denotes a times step. The \textit{left} panel shows the distributions of the posterior samples. The \textit{right} panel shows the times series of the MCMC sampling over 6000 iterations.}
    \label{fig:appendix:pymc}
\end{figure*}

\begin{figure}[H]
    \centering
    \includegraphics[width=0.8\linewidth]{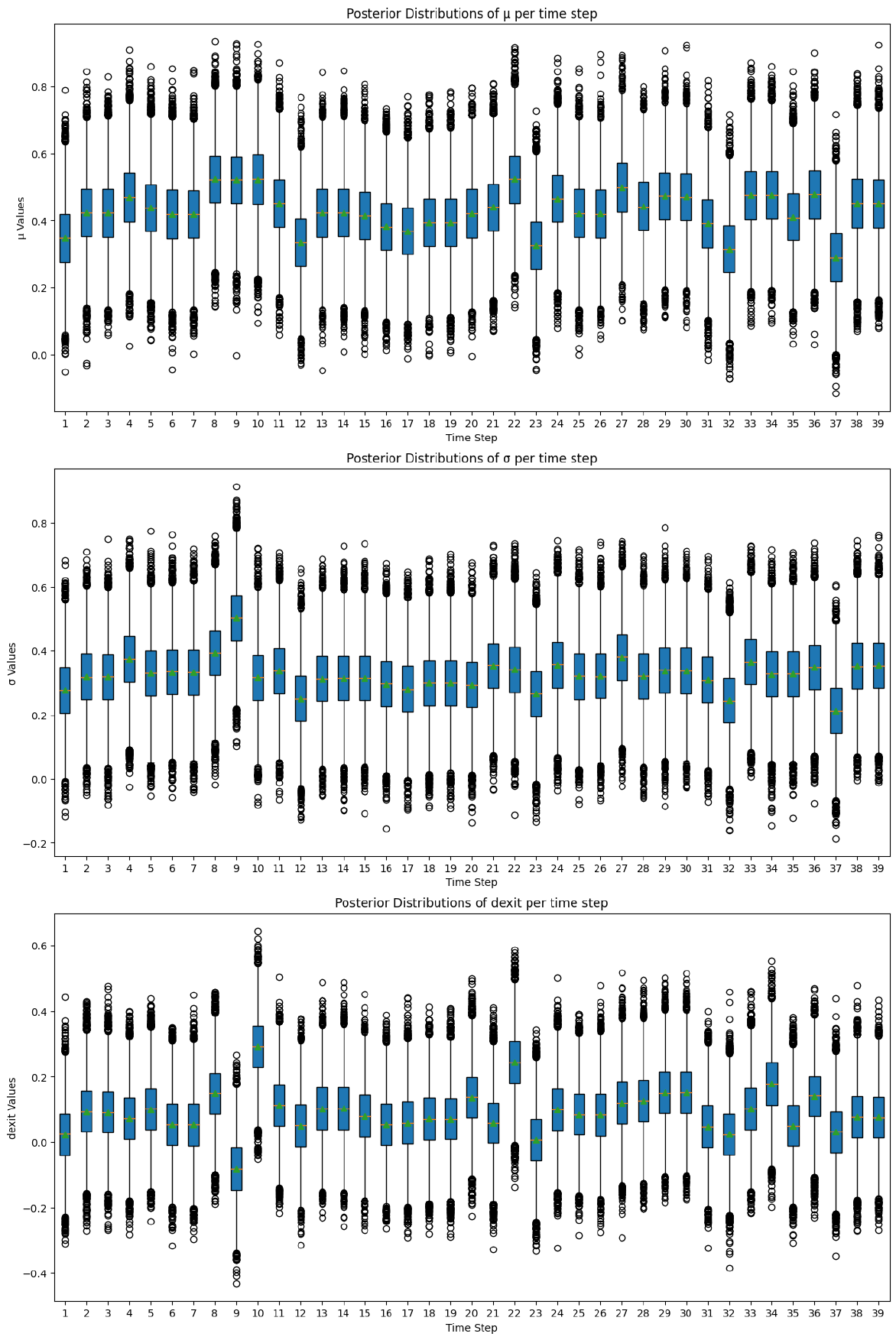}
    \caption{ PyMC posterior parameters for each variable over each time step. Each box-whisker plot denotes a times step.}
    \label{fig:appendix:pymc_2}
\end{figure}

The PyMC Python package \ref{alg:pyMC} generates a trace plot for each of the posterior distributions in the hierarchical Bayesian model detailed in Section \ref{inference}. The package allows the user to visualize the distribution of the posterior samples for each variable individually. The package also plots the sampled values for each variable across the iterations of the background Markov Chain Monte Carlo (MCMC) sampling algorithm. For the hierarchical model in this work, estimates the reaction rate for each each time step. Hence, the posterior distributions are denoted by each individual color. The posterior means of each distribution is used as the reaction rate at time step $t$. Figure~\ref{fig:appendix:pymc} and~\ref{fig:appendix:pymc_2} show to views of the posteriors at each time step.

\newpage

\newpage


\end{document}